\documentclass[11pt,a4paper]{article}
\usepackage{amsmath,amssymb}
\usepackage{epsfig}
\usepackage{axodraw}
\usepackage[sort&compress,square]{natbib}
\usepackage[scale={0.8,0.8},vmarginratio=9:10,hmarginratio=1:1,headheight=20pt,footskip=40pt]{geometry}
\usepackage[dvips]{hyperref}

\newcommand{\ii}{{\rm i}}

\title{
\hfill{\small DCPT/08/176}\\[-0.2cm]\hfill{\small DESY 08-183}\\[-0.2cm]\hfill{\small IPPP/08/88}\\[-0.2cm]\hfill{\small OUTP-08-20P}\\
\vspace{20pt} \Large{\textbf{The Discovery Potential of Laser Polarization Experiments}}}
\author{Markus Ahlers$^{1,}$\footnote{{\bf
e-mail}: m.ahlers1@physics.ox.ac.uk}\,\,, Joerg Jaeckel$^{2,}$\footnote{{\bf
e-mail}: joerg.jaeckel@durham.ac.uk}\,\,, and Andreas Ringwald$^{3,}$\footnote{{\bf
e-mail}: andreas.ringwald@desy.de}
\\[2ex]
\small{\em $^1$Rudolf Peierls Centre for Theoretical Physics, University of Oxford, Oxford OX1 3NP, UK}\\
\small{\em $^2$Institute for Particle Physics and Phenomenology, Durham University, Durham DH1 3LE, UK}\\
\small{\em $^3$Deutsches Elektronen-Synchrotron, Notkestra\ss e 85, 22607 Hamburg, Germany}
}
\date{}

\begin{document}

\maketitle

\begin{abstract}
\noindent
Currently, a number of experiments are searching for vacuum magnetic birefringence and dichroism,
{\it i.e.}~for dispersive and absorptive features in the propagation of polarized light along a
transverse magnetic field in vacuum.
In this note we calculate the Standard Model contributions to these signatures, thereby illuminating the
discovery potential of such experiments in the search for new physics.
We discuss the three main sources for a Standard Model contribution to a dichroism signal: photon splitting, neutrino pair production and production of gravitons.
\end{abstract}

\vspace{3ex}

\section{Motivation}

A number of experiments searching for dispersive and absorptive
features in the propagation of polarized light along a transverse
magnetic field are presently operating
(PVLAS~\cite{Zavattini:2007ee}, Q\&A~\cite{Chen:2006cd}), being
commissioned (BMV~\cite{Rizzo:Patras}), or under serious
consideration (OSQAR~\cite{OSQAR},
PVLAS~Phase~II~\cite{Cantatore:2008ju}). Their primary goal is to verify
the long-standing prediction from quantum electrodynamics (QED) for
these
observables~\cite{Heisenberg:1935qt+,Adler:1970gg,Adler:1971wn,Dittrich:2000zu},
in particular to detect the vacuum magnetic birefringence caused by
virtual electron-positron fluctuations, and to search for possible
contributions of new very weakly interacting sub-eV particles, such
as electrically neutral spin-zero (axion-like)~\cite{Maiani:1986md},
spin-one (photon-like)~\cite{Ahlers:2007rd+}, or
(mini-)charged~\cite{Gies:2006ca+} particles, or other
new low-energy phenomena beyond the Standard Model
({\it e.g.}~\cite{Jaeckel:2006xm+}). Therefore, the leading Standard Model contributions for vacuum
magnetic birefringence and dichroism are of quite some interest.
Since we are mainly interested in observables (nearly) free of a Standard Model background our main focus lies on the dichroism.

The leading QED contribution to vacuum magnetic dichroism,  namely
photon pair production, $\gamma\stackrel{\text{\tiny\it B}}{\to}\gamma\gamma$, also known as
``photon splitting'', has been worked out some time ago and is well
documented~\cite{Adler:1970gg,Adler:1971wn}. Here, we will also
consider the leading contribution from weak interactions, {\it
i.e.}~neutrino pair  production, $\gamma\stackrel{\text{\tiny\it B}}{\to}\nu\bar\nu$, and from
gravitational interactions, {\it i.e.}~the production of gravitons
$\gamma\stackrel{\text{\tiny\it B}}{\to} G$.

\pagebreak[1]
The central questions which we would like to address are:
\begin{itemize}
\item How do the different contributions to the vacuum magnetic dichroism compare to each other?
\item What is the background free discovery potential for new physics in measurements of vacuum magnetic dichroism?
\end{itemize}

To this end, we review in Sect.~\ref{vmdinsm} the photon splitting contribution to the
vacuum magnetic dichroism and calculate the neutrino pair production contribution to the
latter. Moreover, we will estimate the size of the graviton contribution.
We comment on an apparent dichroism caused by birefringence effects in high finesse cavities in Sect.~\ref{FPchapter}.
Finally, in Sect.~\ref{conclusions} we summarize and conclude by giving estimates for the background free
discovery potential for axion-like and minicharged particles.

\section{Vacuum Magnetic Dichroism and Birefringence in the Standard Model}\label{vmdinsm}

We will start with a brief review of the absorptive and dispersive features of light propagating through a magnetic field in vacuum. Specifically, we consider the case of linearly polarized laser light propagating orthogonal to the magnetic field lines. The laser amplitude can be decomposed in components parallel and perpendicular to the magnetic field, $A_\parallel$ and $A_\perp$, respectively.

The dispersion relation $k_i-\omega$ for the two photon polarization states $i=\parallel,\perp$ with momenta $k_i$ and frequency $\omega$ has a real
dispersive ($\propto\Delta n_i$) and an imaginary absorptive ($\propto\ii\kappa_i$) part\footnote{As usual, we work in natural units with $\hbar=c=1$.},
\begin{equation}
k_i - \omega  = \omega\Delta n_i + \frac{\ii\kappa_i}{2}\,,
\end{equation}
compared to the free vacuum propagation with $k-\omega=0$.
The photon-to-photon transition amplitude after a propagation distance $\ell$ can be written as (cf., {\it e.g.}, Refs.~\cite{Ahlers:2007rd+})
\begin{equation}
\label{amplitude}
A^{i}_{\gamma\to\gamma} = \exp(\ii\omega\Delta n_i\ell )\,\exp(-\ell\kappa_i/2)\,.
\end{equation}
From this, the survival probability for an incoming photon
polarization state $i$ can be inferred as
\begin{equation}
P^{i}_{\gamma\to\gamma}=|A^{i}_{\gamma\to\gamma}|^{2}=\exp(-\ell\kappa_i)\approx1-\ell\kappa_i\,,
\end{equation}
Hence, $\ell\kappa_i$ is the photon absorption probability, denoted $\pi_i$ in the following. A linear polarized laser beam entering the magnetic field at an angle
$\theta$ will experience a small rotation
\begin{align}
\label{deltat}
\Delta \theta&=\frac{1}{2}(|A^{\perp}_{\gamma\to\gamma}|-|A^{\parallel}_{\gamma\to\gamma}|)\sin(2\theta)\approx \frac{1}{4}(\pi_\parallel - \pi_\perp)\sin(2\theta)\,,
\end{align}
where the approximation is valid for amplitudes that are close to $1$ and $\kappa_i\ell\ll1$. Phase shifts compared to an unmodified photon beam appear as the
argument of the amplitude,
${\rm{Arg}}(A^{\perp,\parallel}_{\gamma\to\gamma})$.  One finds for
the ellipticity,
\begin{align}
\label{ellipticity}
\psi&=\frac{1}{2}[{\rm{Arg}}(A^{\parallel}_{\gamma\to\gamma})-
{\rm{Arg}}(A^{\perp}_{\gamma\to\gamma})]\sin(2\theta)\approx \frac{1}{2}(\Delta n_\parallel-\Delta n_\perp)\omega\ell\sin(2\theta)\,.
\end{align}
Due to Lorentz invariance neither rotation nor ellipticity appears in the
absence of a magnetic field, and the amplitudes
$A^{\parallel,\perp}_{\gamma\to\gamma}$ are equal.
In the presence of a magnetic field, however, the amplitudes differ,
because the oscillation and absorption lengths are different for
photons parallel $\parallel$ and perpendicular $\perp$ to the magnetic
field.

The contribution to $\Delta n$ and $\psi$ can be enhanced, if one exploits the
possibility to enclose the magnetic field region within an optical cavity where the laser
photons experience many reflections along the optical axis and thus a large number of passes through the magnetic field. In this case Eqs.~(\ref{deltat})
and (\ref{ellipticity}) get an additional factor $N_{\rm pass}\gg 1$ counting the number of reflections.

\begin{figure}[t]
\begin{center}
 \begin{picture}(418,97) (177,-88)
    \SetWidth{0.5}
    \Photon(304,1)(352,1){5.5}{2}
    \ArrowLine(352,1)(416,1)
    \ArrowLine(416,1)(416,-47)
    \ArrowLine(416,-47)(352,-47)
    \ArrowLine(352,-47)(352,1)
    \Photon(352,-47)(352,-79){5.5}{2}
    \Photon(416,-47)(416,-79){5.5}{2}
    \Oval(416,-79)(6.485,6.485)(-45.0)\Line(420.243,-83.243)(411.757,-74.757)\Line(411.757,-83.243)(420.243,-74.757)
    \Oval(352,-79)(6.485,6.485)(-45.0)\Line(356.243,-83.243)(347.757,-74.757)\Line(347.757,-83.243)(356.243,-74.757)
    \Photon(416,1)(464,1){5.5}{2}

    \Text(293,1)[l]{\Large$\gamma$}
    \Text(470,1)[l]{\Large$\gamma$}
    \Text(386,16)[t]{\Large$e$}
    \Text(436,-89)[br]{\large$B$}
    \Text(372,-89)[br]{\large$B$}
  \end{picture}
\end{center}
\caption[]{\small The leading order QED contribution to the photon propagator in an external magnetic field causing vacuum magnetic birefringence.}\label{QEDdiag}
\end{figure}
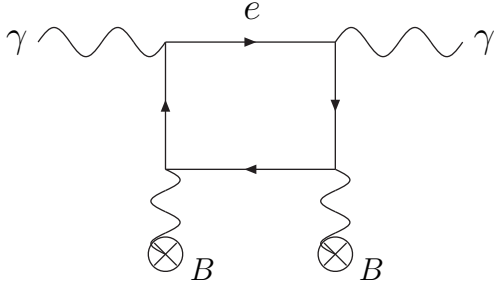

The leading order QED contribution to the refractive index by electron-positron fluctuations in an external magnetic field $B$  (cf.~Fig~\ref{QEDdiag}) is
\begin{equation}
\label{QEDn}
\Delta n_{\parallel,\perp}=\left[\left(7\right)_{\parallel},\left(4\right)_{\perp}\right]\frac{\alpha}{90\pi}\left(\frac{B}{B_{\rm cr}}\right)^2\,,
\end{equation}
where the critical magnetic field is defined as $B_{\rm cr}=m_e^2/e\simeq4.41\times10^9$~T. The corresponding leading order ellipticity induced by QED effects is then
\begin{equation}\label{psiQED}
\psi_{\rm QED} = 1.0\times10^{-17}\bigg(\frac{\omega}{\rm eV}\bigg)\bigg(\frac{\ell}{\rm m}\bigg)\bigg(\frac{B}{\rm T}\bigg)^2N_{\rm pass}\sin(2\theta)\,.
\end{equation}

For optical laser experiments with eV photons the real production of electron-positron pairs is not possible.
The only Standard Model particles which are light enough to contribute to the absorption coefficient are photons, neutrinos and gravitons. We will present the results for the absorption probabilities
$\pi_{\parallel,\perp}$ for photon splitting, neutrino pair production and graviton production in the
following subsections.

\subsection{Photon splitting}

An exhaustive study of photon splitting in an external magnetic field, $\gamma\stackrel{\text{\tiny\it B}}{\to} \gamma\gamma$, has been
conducted by Adler (cf.~Refs.~\cite{Adler:1970gg,Adler:1971wn}).
Taking into account not only absorptive, but also dispersive effects, it was found that, at low energies,
{\it i.e.}~below the electron-positron pair production threshold,
$2 m_e > \omega>0$, the reactions $\perp\to \perp_1 + \perp_2$,
$\parallel \to \perp_1 + \parallel_2$, and $\parallel\to \parallel_1 +\perp_2$ are kinematically
forbidden, leading in particular to
\begin{equation}
\pi_\parallel^{\gamma\gamma} = 0\,,
\end{equation}
while the reaction $\perp\to \parallel_1 + \parallel_2$ occurs, with the
absorption probability, for magnetic field strengths below the critical field
strength, given by~\cite{Adler:1970gg,Adler:1971wn,Dittrich:2000zu}
\begin{equation}
\label{prob_ph_split}
\pi_\perp^{\gamma\gamma} =  \frac{13^2}{3^5\times5^3\times7^2\,\pi^2}\ \alpha^3 \bigg( \frac{B}{B_{\rm cr}}\bigg)^6
\bigg( \frac{\omega}{m_e}\bigg)^5 m_e\, \ell
=4.5\times 10^{-86} \bigg( \frac{B}{\rm T}\bigg)^6
\bigg( \frac{\omega}{\rm eV}\bigg)^5
\bigg( \frac{\ell}{\rm m}\bigg) \,.
\end{equation}
The extreme smallness of the probability of photon splitting results from the fact that diagramma\-ti\-cally
it first appears in the hexagon diagram (cf.~Fig.~3 in Ref.~\cite{Adler:1971wn}), involving
three interactions with the external magnetic field, leading to a suppression by
a factor of $(B/B_{\rm cr})^6$.

\subsection{Neutrino pair production in constant magnetic fields}

In contrast to photon splitting, the leading matrix element for neutrino pair production
involves only one interaction with the external field and is, correspondingly, described
by a box diagram. It can be most easily obtained from the effective
Lagrangian describing processes involving two neutrinos and two photons at energies much below the electron
mass~\cite{Gies:2000wc} (see also Refs.~\cite{Kuznetsov:1997iy+}),
\begin{equation}
\label{leff_nunu}
\mathcal{L}_{\rm eff}^{\nu\bar\nu} =
\frac{G_Fg_A}{\sqrt{2}}
\frac{\alpha}{6\pi}  \frac{1}{m_e^2}
\left\{
-\left( \partial^\mu L_\mu \right)
\left( -\frac{1}{4} F_{\mu\nu} \tilde{F}^{\mu\nu}\right)
+ \left( \partial^\alpha F_{\alpha\beta}\right)
\left( L_\mu \tilde{F}^{\beta\mu}\right)
\right\}
+ {\mathcal O}(1/m_e^4)
\,.
\end{equation}
For simplicity, we start by considering the case of one massive neutrino flavour, either of Dirac ($S=0$) or
Majorana ($S=1$ and $\bar\nu = \nu$) type\footnote{The difference between Dirac and Majorana neutrinos consist
of a factor $(1+S)^{-1}$ in the production probability. The generalization to the more realistic case of three neutrino
flavor and mass eigenstates with non-trivial mixing is considered in Appendix~\ref{numix}. This does not change the order of magnitude of our results.}.
In this case, the neutrino current reads $L_\mu = \bar\nu \gamma_\mu (1+\gamma_5) \nu$.
Note that, in the Standard Model, the axial vector coupling constants are $g_A=1/2$ for $\nu_e$ and
$g_A=-1/2$ for $\nu_{\mu,\tau}$, respectively.

For our case of interest, namely photon-initiated pair production of neutrinos in the
background of a magnetic field, only the first term in Eq.~(\ref{leff_nunu}) contributes
to the matrix element. The differential number of produced neutrino pairs $\nu (p)\bar\nu (p^\prime )$,
with total four momentum $k=p+p^\prime$, is obtained as~\cite{Gies:2000wc}
\begin{equation}
\label{ngnubarnu_Dirac1}
\frac{{\rm d}n_{\nu\bar\nu}}{{\rm d}^4k} =
\frac{1}{1+S}\frac{G_F^2 g_A^2\alpha^2}{9 (2\pi)^7} \frac{m_\nu^2}{m_e^4}
\,\left| \mathcal{G} (k)\right|^2\,
k^2\, \sqrt{1- 4\frac{m_\nu^2}{k^2}}\,
\Theta  \left( k^2- 4m_\nu^2\right)
\,,
\end{equation}
where $\mathcal{G}(k)$ is the Fourier transform
\begin{equation}
\mathcal{G}(k) \equiv \int {\rm d}^4x\ {\rm e}^{{\rm i}k\cdot x}
\left(-\frac{1}{4} F_{\mu\nu} \tilde{F}^{\mu\nu} \right)
= \int {\rm d}^4x\ {\rm e}^{{\rm i}k\cdot x}
 \mathbf E\cdot \mathbf B
\,,
\end{equation}
involving the scalar product between the electric field of the laser beam
and the external magnetic field. The latter is zero, $\mathbf E\cdot \mathbf B = 0$, if the polarization of the
laser beam is perpendicular to the direction of the magnetic field. Correspondingly, the probability
$\pi_\perp$ vanishes,
\begin{equation}
\pi_\perp^{\nu\bar\nu} = 0
\,.
\end{equation}
A non-zero result is obtained, on the other hand, when the
laser beam's polarization is parallel to the magnetic field.

Concretely,
we consider the field configuration
\begin{equation}
\label{eb}
\mathbf{E} =  E\, {\rm e}^{{\rm i}\omega (x-t)}  \mathbf{e}_z, \hspace{6ex}
\mathbf{B} =  B \chi_x ( - \ell/2,+\ell/2 )\mathbf{e}_z
\,,
\end{equation}
where the first equation describes a laser beam propagating in the $x$-direction with
a linear polarization in the $z$ direction and the second equation describes a constant
magnetic field with linear extension $\ell$ in the $x$-direction
($\chi_x(a,b)=1$ for $x\in [a,b]$ and $=0$ otherwise) and pointing in the $z$ direction.
In this set-up one obtains
\begin{equation}\label{Gk}
\mathcal{G}(k) =
EB \ \delta (k_0-\omega ) \delta (k_y) \delta (k_z) (2\pi )^3
\Delta_\ell(k_x-\omega)\,,\quad\text{with}\quad\Delta_\ell(k) = \frac{\sin ( k\ell/2) }{k/2}\,.
\end{equation}
Note that in the limit $\ell\to\infty$ the expression $\Delta_\ell(k)$ reduces to $(2\pi)\delta(k)$.

For the absorption probability of
laser photons polarized along the direction of the magnetic field we find,
\begin{align} \nonumber
\pi_\parallel^{\nu\bar\nu} &=
\frac{1}{1+S}\, \frac{2}{9}
\frac{G_F^2 \alpha^2}{(2\pi)^4} \,\frac{B^2 \omega^4}{m_e^4}\,N_{\rm pass}\
{ I}\left( \frac{m_\nu}{\omega},\omega\ell\right)
\\[1.5ex]
&= 5.76\times 10^{-73} \,\frac{1}{1+S}\,\bigg( \frac{B}{\rm T}\bigg)^2
\bigg( \frac{\omega}{\rm eV}\bigg)^4
 N_{\rm pass}\,\
{ I}\left( \mu,\tau\right)\,,\label{pinunu}
\end{align}
where
\begin{equation}\label{integral}
{ I}(\mu,\tau) = \frac{\mu^2}{4}\!\!\!\int\limits_{-\sqrt{1-4\mu^2}}^{\sqrt{1-4\mu^2}}\!\!\!\!\!\!d\kappa\sqrt{1-\frac{4\mu^2}{1-\kappa^2}}(1-\kappa^2)
|\Delta_\tau(\kappa-1)|^2\,,
\end{equation}
with dimensionless parameters $\mu\equiv m_\nu/\omega$ and $\tau\equiv\ell\omega$.

\begin{figure}[t]
\begin{minipage}[c]{\linewidth}\centering
\includegraphics[width=0.6\linewidth]{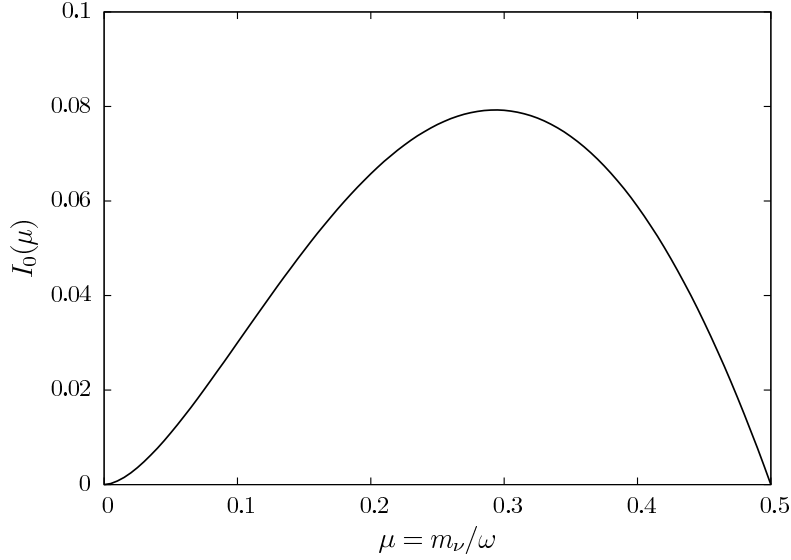}
\end{minipage}
\caption[]{\small The asymptotic value $I_0$ of $I$ for $\tau\to\infty$.}\label{Iintegral}
\end{figure}

In order to compare the probability of pair production~(\ref{pparnubarnu}) with
the one of photon splitting~(\ref{prob_ph_split}), we have to evaluate the parametric
integral~(\ref{integral}).
For all practical purposes, we are interested in its value
for large values of the length $\ell$ in units of the inverse laser energy,
$\tau =5.1\times 10^6\,(\omega/{\rm eV})\,(\ell/{\rm m})\gg 1$.
An integration by parts of Eq.~(\ref{integral}) reveals the following asymptotic expansion
for large $\tau$,
\begin{equation}\label{Iint}
{ I}(\mu,\tau)\, =\, { I}_0(\mu)
 \,+\,\frac{1}{\tau}\times{ J}(\mu,\tau)\,,
\end{equation}
with $|{J}(\mu,\tau)|\leq 0.4$ and
\begin{equation}
\label{largem}
{ I}_0\left( \mu \right) = \frac{\mu^2}{2}\!\!\!\int\limits_{-\sqrt{1-4\mu^2}}^{\sqrt{1-4\mu^2}}\!\!\!\!\!\!d\kappa\sqrt{1-\frac{4\mu^2}{1-\kappa^2}}\frac{1+\kappa}{1-\kappa}\,.
\end{equation}
The dependence of $I_0$ on the relative neutrino mass $\mu$ is shown in Fig.~\ref{Iintegral}. The evident bound $I_0(\mu)<0.08$ translates
into a bound on the absorption coefficient from neutrino
pair production in the limit $\tau\gg1$ as
\begin{align}
\label{pparnubarnu}
\pi_\parallel^{\nu\bar\nu}\leq 4.6\times 10^{-74} \bigg( \frac{B}{\rm T}\bigg)^2
\bigg( \frac{\omega}{\rm eV}\bigg)^4
 N_{\rm pass}\,.
\end{align}

\subsection{Neutrino pair production in alternating magnetic fields}

The production of massive particles from massless laser photons requires a momentum contribution from the background magnetic field.
Accordingly, the production probability is suppressed in more or less constant magnetic fields. Laser experiments can extend their sensitivity by considering alternating
magnetic fields\footnote{This has been noted in the context of axion-like particles~\cite{VanBibber:1987rq,Raffelt:1987im}. An alternative for axion-like particles is to insert phase-shift plates as proposed in~\cite{Jaeckel:2007gk}, however, in laser polarization experiments this might be experimentally more challenging.}.
For simplicity, we consider a sinusoidal magnetic field of the form
\begin{equation}
\mathbf{B}(x,t) = B\,\mathbf{e}_z\,\chi_x ( - \ell/2,+\ell/2 )\times\begin{cases}\cos\left(\frac{N\pi}{\ell}x\right)&\text{for $N$ even}\\\sin\left(\frac{N\pi}{\ell}x\right)&\text{for $N$ odd}\end{cases}\,,
\end{equation}
where $N$ counts the number of field alternations. In particular, these alternating magnetic fields are present as undulators in free electron lasers generating keV laser photons~\cite{FEL}.
In this set-up one obtains
\begin{equation}\label{deltaalt}
\left|\Delta_{\ell,N}(k)\right| = \left|\frac{2k}{k^2-k_\text{res}^2}\right|\times\begin{cases}\sin\left(\frac{k\ell}{2}\right)&\text{for $N$ even}\\\cos\left(\frac{k\ell}{2}\right)&\text{for $N$ odd}\end{cases}\,,
\end{equation}
with $k_{\rm res} \equiv N\pi/\ell$. In realistic experiments we consider the situation with a fixed width $d$ of the magnetic domains much smaller than $\ell=(N+1)d$. In the limit of large $N$ we can approximate expression (\ref{deltaalt}) by its limit
\begin{equation}
\lim_{N\to\infty}|\Delta_{\ell,N}(k)| = \pi\left|\delta(k-k_{\rm res})+(-1)^N\delta(k+k_{\rm res})\right|\,,
\end{equation}
with $k_{\rm res} = \pi/d \simeq 6.2\times10^{-5}\times({\rm cm}/d)\,{\rm eV}$. For the production of on-shell neutrinos only the second term $\propto\delta(k+k_{\rm res})$ contributes. Energy-momentum conservation requires $2\omega k_{\rm res}-k_{\rm res}^2> 4m_\nu^2$. For typical neutrino masses of the order of $0.1$~eV and free electron laser experiments with $10$~keV photons and $1$~cm undulators we can write the absorption probability as
\begin{align}\nonumber
\pi_\parallel^{\nu\bar\nu} &\simeq
 \frac{1}{1+S}
\frac{G_F^2 \alpha^2}{36(2\pi)^3} \frac{m_\nu^2 B^2}{m_e^4}\,\omega^2 k_{\rm res}\,\ell\,N_{\rm pass}\qquad\qquad\qquad\text{for}\quad \omega\gg k_{\rm res}\gg2m_\nu^2/\omega\\ \label{neutrinomax}
&\simeq 1.4\times 10^{-63} \frac{1}{1+S}\bigg( \frac{m_\nu}{0.1{\rm eV}}\bigg)^2\bigg( \frac{B}{\rm T}\bigg)^2
\bigg( \frac{\omega}{10\,{\rm keV}}\bigg)^{2}\bigg( \frac{d}{\rm cm}\bigg)^{-1}\bigg( \frac{\ell}{10 {\rm m}}\bigg)
 N_{\rm pass}\,.
\end{align}
For the indicated experimental benchmark the neutrino production probability is increased by about a factor $100$ compared to the constant magnetic field case (Eq.~(\ref{pinunu})).

\subsection{Contribution from graviton production}

\begin{figure}[t]
\begin{center}
  \begin{picture}(378,97) (47,-88)
    \SetWidth{0.5}
    \Photon(48,1)(128,1){5.5}{4}
    \Photon(128,1)(128,-79){5.5}{4}
    \Gluon(128,1)(208,1){5.5}{4}
    \Oval(128,-79)(6.485,6.485)(-45.0)\Line(132.243,-83.243)(123.757,-74.757)\Line(123.757,-83.243)(132.243,-74.757)

    \Text(37,1)[l]{\Large$\gamma$}
    \Text(212,1)[l]{\large$G$}
    \Text(148,-89)[br]{\large$B$}

    \SetOffset(-40,0)
    \Photon(304,1)(352,1){5.5}{2}
    \Gluon(416,1)(464,1){5.5}{2}
    \ArrowLine(352,1)(416,1)
    \ArrowLine(416,1)(416,-47)
    \ArrowLine(416,-47)(352,-47)
    \ArrowLine(352,-47)(352,1)
    \Photon(384,-47)(384,-79){5.5}{2}
    \Photon(352,-47)(352,-79){5.5}{2}
    \Photon(416,-47)(416,-79){5.5}{2}
    \Oval(416,-79)(6.485,6.485)(-45.0)\Line(420.243,-83.243)(411.757,-74.757)\Line(411.757,-83.243)(420.243,-74.757)
    \Oval(384,-79)(6.485,6.485)(-45.0)\Line(388.243,-83.243)(379.757,-74.757)\Line(379.757,-83.243)(388.243,-74.757)
    \Oval(352,-79)(6.485,6.485)(-45.0)\Line(356.243,-83.243)(347.757,-74.757)\Line(347.757,-83.243)(356.243,-74.757)

    \Text(293,1)[l]{\Large$\gamma$}
    \Text(470,1)[l]{\large$G$}
    \Text(386,16)[t]{\Large$e$}
    \Text(436,-89)[br]{\large$B$}
    \Text(404,-89)[br]{\large$B$}
    \Text(372,-89)[br]{\large$B$}

  \end{picture}
\end{center}
\caption[]{\small Leading order contributions to the graviton production in an external magnetic field.}\label{gravitondiag}
\end{figure}
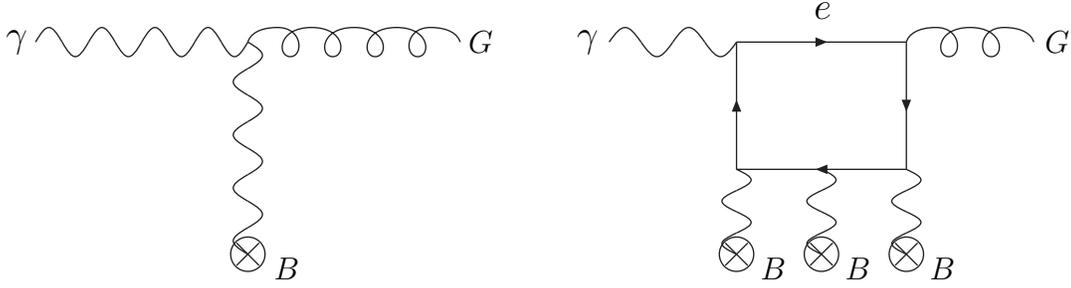

The only remaining particle in the Standard Model that is light enough to be produced in a laser experiment is the graviton.
In the presence of a magnetic field gravitons can mix with photons in a way similar to axions~\cite{Raffelt:1987im} (cf.~also Fig.~\ref{gravitondiag}).
There are, however, two crucial differences. The first is simply that gravitons are completely massless. The second is that while axions
couple only to one polarization gravitons couple to both polarizations with equal strength. Accordingly, to leading order $\pi_{\parallel}=\pi_{\perp}$
and the leading order contribution to a rotation of the laser polarization (cf.~Eq.~\eqref{deltat}) cancels.

The QED vacuum birefringence (see Fig.~\ref{QEDdiag}), however, leads to a small difference in the graviton production for the different polarizations. The reason is that
the QED effect causes a phase difference between the photon and the graviton. In consequence, gravitons at
different positions along the beam line have different phases and constructive interference is disturbed\footnote{This effect is analogous to the
effect of the axion mass in the axion-photon system.}.
Since the phase difference caused by QED depends on the polarization (this is exactly the birefringence property) graviton production, too, depends on the polarization
and we can observe a dichroism.
Moreover, at one-loop order (cf.~right hand side of Fig.~\ref{gravitondiag}) there is also a difference in the photon-graviton amplitude between the two different
polarizations.

Following~\cite{Raffelt:1987im} and including also the one-loop (electrons in the loop) correction to the photon-graviton amplitude~\cite{Bastianelli:2007jv} we
find
\begin{equation}
\pi^{\rm graviton}_{\parallel,\perp}= 2\left(\frac{B}{M_{\rm Pl}}\right)^2
\left(1+C_{\parallel,\perp}\frac{\alpha}{4\pi}\left(\frac{B}{B_{\rm cr}}\right)^{2}\left(\frac{\omega}{m_{e}}\right)^{2}\right)^2
\left(\frac{1}{\omega \Delta n_{\parallel,\perp}}\right)^2
\sin^{2}\left(\frac{\omega \Delta n_{\parallel,\perp}\ell}{2}\right),
\end{equation}
where we have used the QED contribution to the refractive index given in Eq.~(\ref{QEDn}) and the coefficients
\begin{equation}
C_{\parallel}=\frac{10}{63}\,,\qquad C_{\perp}=\frac{32}{315}\,.
\end{equation}
Expanding the rotation $\Delta
\theta=(\pi_{\parallel}-\pi_{\perp})/4$ in powers of $B/B_{\rm cr}$ the
leading 0th order terms cancel. At next order we find,
\begin{align}
\label{gravitoncontrib}\nonumber
\Delta \theta_{\rm graviton}&=\left[\frac{1}{140}\bigg(\frac{\alpha}{2\pi}\bigg)^{2}\left(\frac{B}{B_{\rm cr}}\right)^2\left(\frac{\omega}{m_{e}}\right)^{2}
-\frac{11}{128}\bigg(\frac{\alpha}{45\pi}\bigg)^{2}\bigg(\frac{B}{B_{\rm cr}}\bigg)^4
(\omega\ell)^2\right]\bigg(\frac{B\ell}{M_{\rm Pl}}\bigg)^{2} N_{\rm pass}
\\
&\simeq\left[3.1\times10^{-76}\bigg(\frac{B}{\rm T}\bigg)^{4}\bigg(\frac{\ell}{\rm m}\bigg)^{2}\bigg(\frac{\omega}{\rm eV}\bigg)^{2}
-2.6\times10^{-72}\bigg(\frac{B}{\rm T}\bigg)^6\bigg(\frac{\ell}{\rm m}\bigg)^{4}\bigg(\frac{\omega}{\rm eV}\bigg)^{2}\right]N_{\rm pass}\,.
\end{align}

In Tab.~\ref{table1} we have listed the total rotation expected from
Standard Model contributions for various experimental setups. It is
interesting to note that for the parameters of these experiments the
largest contribution arises from graviton production, a somewhat
smaller one from the neutrinos and the smallest is photon splitting.

\begin{table}[t]\centering
\renewcommand{\arraystretch}{2.0}\small
\begin{tabular}{l|cccc|cc}\hline
Experiment&$\omega$[eV]&$\mathcal{F}_{\rm max}$&$B_{\rm{max}}$[T]&$\ell_B$[m]&$|\Delta \theta_{\rm SM}|$~[nrad]&$|\psi_{\rm SM}|$~[nrad]\\
\hline
BMV&$1.17$&200000&$12.3$&$0.365$&$2.8\times10^{-53}$&$8.2\times10^{-2}$\\
OSQAR&$1.17$&10000&$9.5$&$14.3$&$7.0\times10^{-49}$&$9.6\times10^{-2}$\\
PVLAS&$2.33$&70000&$5.0$&$1.0$&$9.8\times10^{-54}$&$2.6\times10^{-2}$\\
PVLAS II&$2.33$&220000&$2.3$&$0.5$&$1.8\times10^{-56}$&$8.6\times10^{-3}$\\
Q\&A&$1.17$&$31000$&$2.3$&$0.6$&$1.3\times10^{-57}$&$7.3\times10^{-4}$\\
\hline
\end{tabular}
\caption[]{\small Parameters of the polarization experiments searching for a possible rotation
$\Delta \theta$ of the polarization after passage
through a magnetic field. $\omega$ is the frequency of the laser light, $\mathcal{F}_{\rm max}$ is the maximal finesse of the cavity containing the magnetic field of maximal strength $B_{\rm max}$ and length $\ell_{B}$. The Standard Model contribution to the rotation, $\Delta\theta_{\rm SM}$, is dominated by graviton production (Eq.~(\ref{gravitoncontrib})).
}\label{table1}
\end{table}

\section{Effects of the Experimental Apparatus\label{FPchapter}}

There are additional contributions to the vacuum magnetic dichroism
originating from the experimental apparatus. Most experiments increase
their sensitivities by the extension of the optical path in mirror
systems, often by using Fabry-P\'erot cavities. High finesse cavities posses resonant eigenmodes
with extremely narrow line widths (see Appendix \ref{FPapp}). In the presence of magnetic birefringence -- either as an effect of a non-ideal apparatus or a true signal --
the wavelengths of the orthogonal and parallel modes are different.
Consequently, the two polarizations cannot simultaneously be in perfect resonance with the cavity.
As a result, at least one mode is attenuated. This can induce a rotation of the laser polarization plane~\cite{Zavattini2006}.

As an example, let us consider an idealized situation where the only source of birefringence is the QED effect (cf.~Eq.~(\ref{QEDn})). In the presence of vacuum magnetic birefringence the laser propagation modes orthogonal and parallel to the external magnetic field have $k_{\perp,\parallel} \approx \omega(1 + \Delta n_{\perp,\parallel})$.
Using Eqs.~\eqref{deltat}, \eqref{transmission} and \eqref{finesse} we find that to leading order a Fabry-P\'erot cavity which is locked
to a resonance $\omega\ell=\mathbf{N}\pi$ will produce a rotation
\begin{align}\label{thetaFP}
\Delta\theta_{\rm FP}&= \frac{1}{2}\frac{2R}{(1-R)^2}(\Delta n^2_\perp - \Delta n^2_\parallel)(\omega\ell)^2\sin(2\theta)\approx\frac{\mathcal F^2}{\pi^2}(\Delta n^2_\perp - \Delta n^2_\parallel)(\omega\ell)^2\sin(2\theta)
\\\nonumber
&\simeq-1.5\times10^{-24}\bigg(\frac{{\cal F}}{10^5}\bigg)^2\bigg(\frac{B}{\text{T}}\bigg)^4\bigg(\frac{\ell}{\text{m}}\bigg)^2\bigg(\frac{\omega}{\text{eV}}\bigg)^2\sin(2\theta)\,,
\end{align}
where $\mathcal{F}$ is the finesse of the cavity (cf.~Appendix \ref{FPapp}) and we have used the QED contribution of Eq.~(\ref{QEDn}).

Already this contribution of QED effects can be huge compared to the Standard Model backgrounds discussed in the previous Sect.~\ref{vmdinsm}. For typical experimental values we find
\begin{equation}\label{FP}
\left|\Delta \theta_\text{FP, QED}\right|\sim \bigg(\frac{{\cal F}}{10^5}\bigg)^{2}\,(10^{-14}-10^{-8})\,{\rm nrad}\,.
\end{equation}
The small magnetic birefringence induced by QED effects is in general not the dominant contribution in realistic experiments.
More importantly, reflections at the cavity mirrors can introduce a small phase shift $10^{-7}$ -- $10^{-6}$ rad per pass between
the orthogonal and parallel field component, that accumulates with the number of reflections $N_{\rm pass}$~\cite{Zavattini2006}. In realistic cavities this is the dominant effect\footnote{
Residual gases in the cavity vacuum might become birefringent in the presence of the magnetic field, the \emph{Cotton-Mouton effect}, however, this
effect is sub-dominant in high quality vacua.}.

However, it should be noted that this effect is
an artifact of a specific experimental setup which includes a cavity. Moreover, even in setups with cavities it has been shown that the crosstalk between birefringence and dichroism can be eliminated by a suitable experimental technique~\cite{Zavattini2006}. In short, the size of the rotation depends on the way in which the cavity is locked to the laser.
For example, if the laser is locked to the cavity with a condition $\omega\ell(1+\Delta n_{0})=\mathbf{N}\pi$ with $\Delta n_{0}\sim \Delta n_{\parallel,\perp}\ll 1$
we find for the rotation,
\begin{equation}
\Delta\theta_{\rm FP}=\frac{\mathcal F^2}{\pi^2}\left[(\Delta n_\perp-\Delta n_{0})^2 - (\Delta n_\parallel-\Delta n_{0})^2\right](\omega\ell)^2\sin(2\theta).
\end{equation}
Adjusting $\Delta n_{0}$ one can change the rotation.
Using this and heterodyne detection, where different birefringence sources show up in different frequency sidebands, one can disentangle or eliminate the rotation caused by the cavity~\cite{Zavattini2006}.

\section{Conclusions}\label{conclusions}

In these notes we have discussed the various Standard Model sources contributing to a rotation of the polarization of light
when passing through a magnetic field.
For typical experimental setups with magnetic fields of the order of $1$\,T, length of the order of $1$\,m and frequencies in the $1$\,eV range we find
that the largest contribution arises from the production of gravitons (cf.~Eq.~\eqref{gravitoncontrib}), a somewhat smaller contribution
arises from neutrino pair production (cf.~Eqs.~\eqref{pparnubarnu}, \eqref{neutrinomax}) and the smallest
contribution is photon splitting (cf.~Eq.~\eqref{prob_ph_split}). Table~\ref{table1} shows the leading order contributions to the Standard Model rotation $\Delta\theta_{\rm SM}$ and ellipticity $\psi_{\rm SM}$ for various laser polarization experiments. Allowing for some room in the experimental parameters the expected order of magnitude for
the rotation lies in the range
\begin{equation}
\label{SM}
\frac{\left|\Delta\theta_\text{SM}\right|}{N_{\rm pass}} \sim (10^{-62}-10^{-51})\,{\rm nrad}\,.
\end{equation}

Comparing this (cf.~Eq.~(\ref{SM})) to the expected rotation from an axion-like particles (ALP) (in the limit of vanishing mass),
\begin{equation}
\frac{\left|\Delta \theta_\text{ALP}\right|}{N_{\rm pass}} \sim \frac{B^{2}\ell^2}{M^{2}_{a}}\,,
\end{equation}
the Standard Model background is negligible even for axion scales $M_a$ way beyond the Planck scale.
Similarly, minicharged particles (MCP) (again in the limit of vanishing mass) lead to a rotation,
\begin{equation}
\frac{\left|\Delta\theta_\text{MCP}\right|}{N_{\rm pass}}\sim 2.1\times10^4
\epsilon^{\frac{8}{3}}\bigg(\frac{B}{\text{T}}\bigg)^{\frac{2}{3}}\bigg(\frac{\ell}{\text{m}}\bigg)\bigg(\frac{\omega}{\text{eV}}\bigg)^{-\frac{1}{3}}\,.
\end{equation}
Accordingly, the background free
discovery potential is in the range $\epsilon\sim 10^{-28}-10^{-25}$ which is very promising when compared to typical predicted values, {\it e.g.},
in realistic string compactifications, ranging from $10^{-16}$ to $10^{-2}$ \cite{Dienes:1996zr}.

In addition to these true Standard Model backgrounds to a dichroism causing a rotation of the laser light, realistic experiments can have additional experimental backgrounds. For example, in Fabry-P\'erot cavities there is a crosstalk between birefringence and dichroism. However, as demonstrated in~\cite{Zavattini2006} these effects can be controlled by suitable measurement techniques.

Overall, optical measurements of vacuum magnetic dichroism allow for an enormous discovery potential for new physics untainted by Standard Model backgrounds, motivating further experimental efforts to go beyond the current sensitivities.

\section*{Acknowledgments}

The authors would like to thank Giovanni Cantatore, Javier Redondo
and Giuseppe Ruoso for helpful discussions, suggestions and comments. MA
acknowledges support by STFC UK (PP/D00036X/1).

\begin{appendix}
\section{Neutrino Pair Production and Flavor Mixing}\label{numix}

The neutrino flavor eigenstates $\nu_\alpha$ ($\alpha=e,\mu,\tau$) are composed out of mass eigenstates $\nu_i$ ($i=1,2,3$) according to
\begin{equation}
\nu_\alpha = \sum_iU^*_{\alpha i}\nu_i\quad\text{and}\quad \nu_i = \sum_\alpha U_{\alpha i}\nu_\alpha\,\,,
\end{equation}
with $U^\dagger U=\mathbf{1}$. The neutrino currents of flavor $\alpha$ are generalized to
\begin{equation}
L^\mu_\alpha = \sum_{i,j} U_{\alpha i}U^*_{\alpha j}L^\mu_{ij}\quad\text{and}\quad L^\mu_{ij} = \sum_\alpha U^*_{\alpha i}U_{\alpha j}L^\mu_\alpha\,\,.
\end{equation}
Note, that for on-shell neutrino mass eigenstates $\nu_i$ with mass $m_i$ we have now vector \emph{and} axial-vector contributions in the interaction term Eq.~(\ref{leff_nunu}) according to mass splittings, {\it i.e.},
\begin{align}
\partial_\mu L^\mu_{\alpha}=\sum_{ij}U_{\alpha i}U^*_{\alpha j} \partial_\mu L^\mu_{ij}=\sum_{ij}U_{\alpha i}U^*_{\alpha j}\left[i(m_i-m_j)\bar\nu_i\nu_j+i(m_i+m_j)\bar\nu_i\gamma_5\nu_j\right]\,.
\end{align}
Since $g^{e}_A=\frac{1}{2}$, but $g^{\mu,\tau}_A=-\frac{1}{2}$ the sum over intermediate flavor eigenstates $\nu_\alpha$ in the process $\gamma\stackrel{\text{\tiny\it B}}{\to}\bar\nu_i\nu_j$ does not simply reduce to mass diagonal terms. Instead, the neutrino production probability of Eq.~(\ref{pinunu}) (summed over all mass eigenstates) is generalized by the substitution
\begin{equation}
3g_A^2I(\mu,\tau)\to\sum_{i,j} \left|\frac{1}{2}\delta_{ij} - U_{e i}U^*_{e j}\right|^2I(\mu_i,\mu_j,\tau)\,.
\end{equation}
The generalized parametric integral (Eq.~(\ref{integral})) of the transition probability is
\begin{equation}
{ I}(\mu_i,\mu_j,\tau) = \frac{\mu_i\mu_j}{4}\!\!\!\!\int\limits_{-\sqrt{1-(\mu_i+\mu_j)^2}}^{\sqrt{1-(\mu_i+\mu_j)^2}}\!\!\!\!\!\!\!d\kappa\,\,\sqrt{1-\frac{2(\mu_i^2+\mu_j^2)}{1-\kappa^2}+\left(\frac{\mu_i^2-\mu_j^2}{1-\kappa^2}\right)^2}(1-\kappa^2) |\Delta_\tau(\kappa-1)|^2\,.
\end{equation}

\section{High Finesse Fabry-P\'erot Cavities}\label{FPapp}

\begin{figure}[t]
\begin{center}
\includegraphics[width=0.4\linewidth]{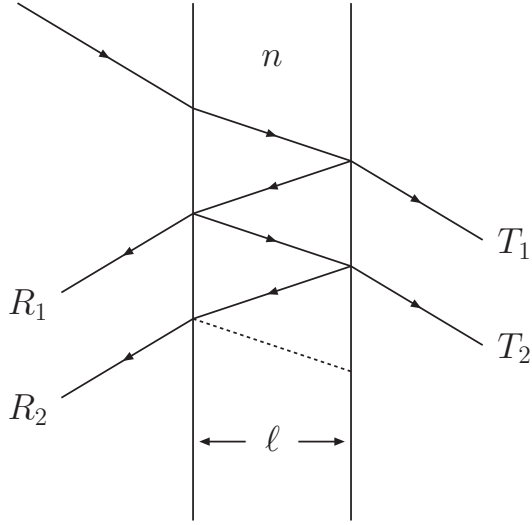}
\end{center}
\caption{\small Light path inside a Fabry-P\'erot cavity.}
\label{fabry}
\end{figure}

Consider a Fabry-P\'erot cavity as sketched in Fig.~\ref{fabry}. The laser light (for illustration shown with a small incident angle) is transmitted and reflected by both cavity mirrors with coefficients $T$ and $R=1-T$, respectively. The transmitted laser amplitude $T_n$ (relative to the initial amplitude right before the first mirror at $z=0$) is
\begin{equation}
T_n = T\,\exp(-{\rm i} k\ell)\,R^{n-1}\,\exp(-{\rm i}2(n-1)k\ell)\,.
\end{equation}
The sum of these amplitudes gives
\begin{equation}
A_\gamma = T_1+\ldots+T_N =  T\,\exp(-{\rm i} k\ell)\,\frac{1-\exp(-{\rm i}2Nk\ell)R^N}{1-\exp(-{\rm i}2k\ell)R}\xrightarrow{N\to\infty}\frac{T\,\exp(-{\rm i} k\ell)}{1-\exp(-{\rm i}2k\ell)R}\,.
\end{equation}
The transmission coefficient of the Fabry-P\'erot cavity is then given by
\begin{equation}
\label{transmission}
|A_\gamma|^2 = T^2\frac{1+R^{2N}-2R^N\cos(2Nk\ell)}{1+R^2-2R\cos(2k\ell)} \xrightarrow{N\to\infty}\frac{T^2}{1+R^2-2R\cos(2k\ell)}\,.
\end{equation}
Hence, the resonant modes of the cavity are $k\ell = \mathbf{N}\pi$.
The \emph{finesse} of the cavity is defined as the ratio of the
spread of the eigenmodes $\Delta \lambda$ over the line width
$\delta\lambda$,
\begin{equation}
\label{finesse}
{\mathcal F} = \frac{\Delta\lambda}{\delta\lambda} = \frac{\pi}{2\arcsin\left(\frac{1-R}{2\sqrt{R}}\right)}\approx
\frac{\pi\sqrt{R}}{1-R}\,.
\end{equation}
In the last expression we have used $1-R\ll1$. The effective number of passes of the laser photons between the cavity mirrors can be estimated from the finesse as
\begin{equation}
N_{\rm pass} \approx \frac{2}{\pi}\mathcal{F}\,.
\end{equation}
\end{appendix}

\end{document}